\documentclass[preprint]{revtex4-2}
\usepackage{color}
\usepackage{graphicx}
\usepackage{amssymb}
\usepackage{amsmath}
\usepackage{todonotes}
\usepackage[hidelinks=true]{hyperref}
\hypersetup{
		colorlinks = true,
		urlcolor = blue,
		linkcolor = blue,
		citecolor = blue
	}

\begin{document}

\title{Thickness-dependent, tunable anomalous Hall effect in hydrogen-reduced PdCoO$_2$ thin films}

\author{Gaurab Rimal}
\email{gr380@physics.rutgers.edu}
\affiliation{Department of Physics \& Astronomy, Rutgers, The State University of New Jersey, Piscataway, New Jersey 08854, USA }

\author{Yiting Liu}
\affiliation{Department of Physics \& Astronomy, Rutgers, The State University of New Jersey, Piscataway, New Jersey 08854, USA }
\affiliation{State Key Laboratory of Precision Spectroscopy, East China Normal University, Shanghai 200062, China }


\author{Matthew Brahlek}
\affiliation{Materials Science and Technology Division, Oak Ridge National Laboratory, Oak Ridge, Tennessee 37831, USA }


\author{Seongshik Oh}
\email{ohsean@physics.rutgers.edu}
\affiliation{Department of Physics \& Astronomy, Rutgers, The State University of New Jersey, Piscataway, New Jersey 08854, USA }
\affiliation{Center for Quantum Materials Synthesis, Rutgers, The State University of New Jersey, Piscataway, New Jersey 08854, USA }

\newcommand{\red}[1]{\textcolor{red}{#1}}
\newcommand{\blue}[1]{\textcolor{blue}{#1}}
\newcommand{\green}[1]{\textcolor{green}{#1}}

\begin{abstract}

	It was recently reported that hydrogen-reduced PdCoO$_2$ films exhibit strong perpendicular magnetic anisotropy (PMA) with sign tunable anomalous Hall effect (AHE). Here, we provide extensive thickness-dependent study of this system, and show that the electronic and magnetic properties are strongly dependent on the thickness and annealing conditions. Below a critical thickness of 25 nm, AHE shows clear PMA with hysteresis, and its sign changes from positive to negative, and back to positive as the annealing temperature increases from 100 $^\circ$C to 400 $^\circ$C. Beyond the critical thickness, both PMA and AHE hysteresis disappear and the AHE sign remains positive regardless of the annealing parameters. Our results show that PMA may have a large role on AHE sign-tunability and that below the critical thickness, competition between different AHE mechanisms drives this sign change. 

\end{abstract}

\maketitle

\section*{Introduction}

Ferromagnetic materials exhibit anomalous Hall effect (AHE), which is an extra contribution added to the ordinary Hall effect and is usually proportional to their magnetization (thus frequently hysteretic). One of the common characters of any AHE is its well-defined sign: depending on how its magnetization couples with the mobile carriers, AHE can be either positive or negative against the applied external magnetic field. In clean materials, the net Berry curvature, integrated at the Fermi level of the materials band structure, determines the sign of AHE. However, in materials with structural disorders or impurities, both magnitude and sign of AHE can be dominated by extrinsic factors such as defect scattering.  Regardless, in most ferromagnetic materials, the AHE sign is not switchable, but certain materials such as SrRuO$_3$ \cite{Haham2011}, chiral magnets \cite{Nakatsuji2015,Kiyohara2016,Nayak2016}, and magnetic topological insulators \cite{Zhang2020,Wang2021}, can exhibit AHE sign reversal depending on external stimulations. 


The non-magnetic, metallic delafossites PtCoO$_2$ and PdCoO$_2$ are the most conducting oxides, with PdCoO$_2$ having the longest mean free path of 20 $\mu$m among oxides \cite{Hicks2012,Mackenzie2017}. Although PdCoO$_2$ is not magnetic, it exhibits itinerant surface magnetism \cite{Mazzola2017,Mazzola2022,Harada2020}, and recent work on hydrogenated PdCoO$_2$ reported the emergence of bulk magnetism with strong perpendicular magnetic anisotropy (PMA). It was shown that these hydrogenated films exhibit  sign change of AHE with hydrogenation parameters \cite{Rimal2021,Liu2022}, likely due to structural changes when PdCoO$_2$ is reduced. Here, we expand on this previous study with comprehensive experiments on over 80 hydrogenated PdCoO$_2$ samples of various thickness and find that as the film gets thicker, PMA gradually turns into in-plane magnetic anisotropy (IMA) and the sign-tunability of AHE also disappears. 

\section*{Experimental details}

We grew PdCoO$_2$ films on Al$_2$O$_3$ (0001) substrates using oxygen plasma assisted molecular beam epitaxy (MBE) \cite{Brahlek2019}. The films were grown at 300 $^\circ$C in a background pressure of $4 \times 10^{-6}$ Torr plasma oxygen in a layer-by-layer fashion. Plasma is generated using a 13.6 MHz RF source at a power of 450 W. After the growth, hydrogenation is carried out at ambient pressure by annealing in a 10\% H$_2$/90\% Ar mixture gas at various temperatures for different duration. All thickness values correspond to the nominal PdCoO$_2$ thickness. Transport measurements were carried out using standard DC van der Pauw technique.


\begin{figure}[ht]
	\centering
	\includegraphics[width=0.95\linewidth]{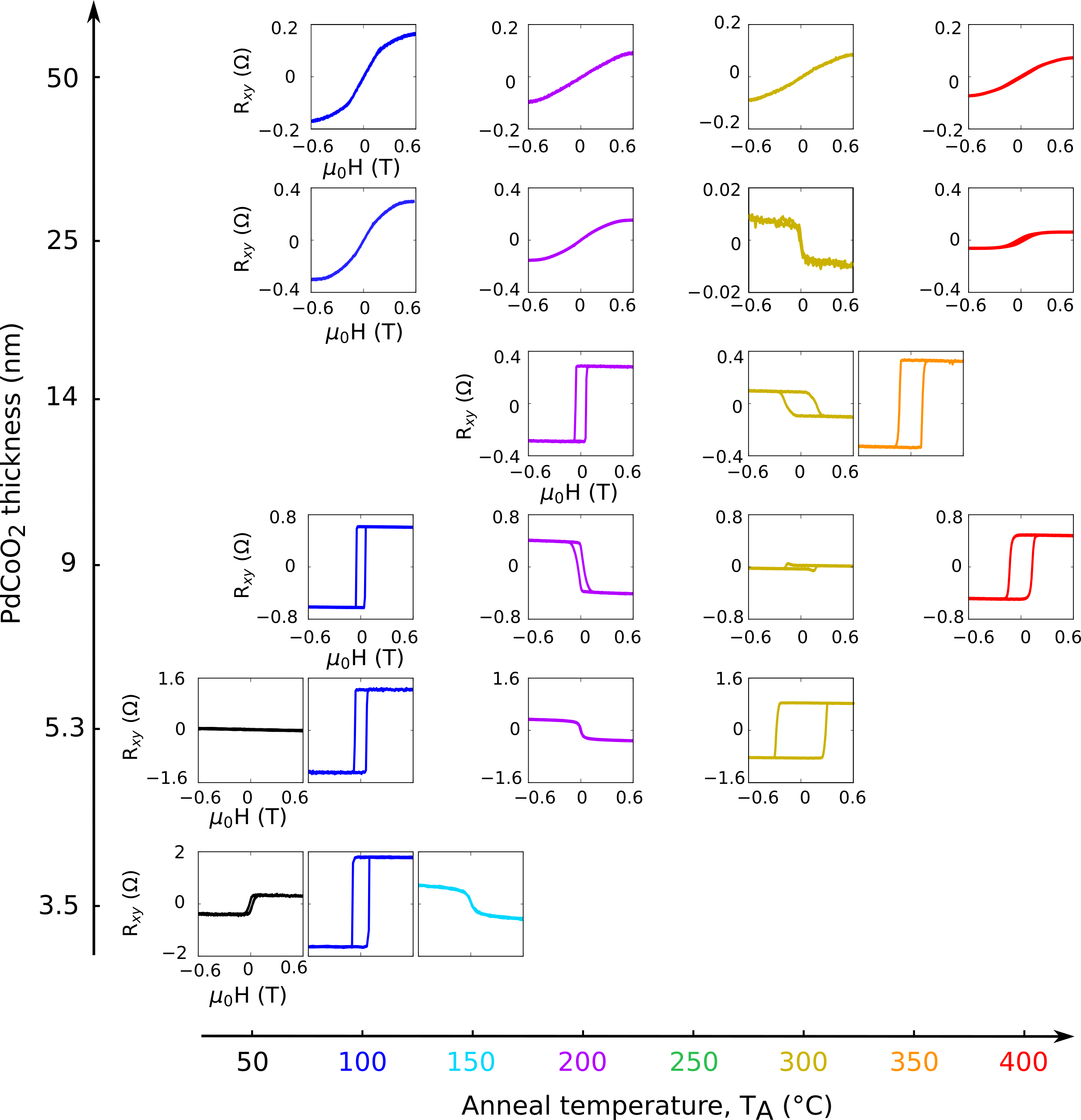}
	\caption{AHE at 295 K for films of different thickness processed at different anneal temperatures for 30 mins. The magnetic behavior of the films is dependent on both the thickness and anneal temperature, which is demonstrated by the change in the overall shape of AHE. } 
	\label{fig1}
\end{figure}

\section*{Results and discussion}

In Figure 1, we demonstrate the effects of anneal temperature and film thickness on the magnetic and transport properties of the PdCoO$_2$ films. Previously, annealing time ($t_A$) and temperature ($T_A$) were used to switch the AHE sign \cite{Rimal2021}. Here we use T$_A$ as the main tuning parameter. Except for the thinnest film (3.5 nm), no sign change occurs for T$_A$ $<$ 100 $^\circ$C. With slightly higher T$_A$ ($\sim$100 - 200 $^\circ$C), AHE with positive sign appears (positive AHE is defined as R$_{xy}^{sat} > 0$ for H$ > 0$). With higher T$_A$, positive AHE abruptly switches to negative in films with thickness up to 25 nm. At even higher anneal temperatures, the AHE sign reverses back to positive. Robust PMA with well-defined coercive fields is present in films with thickness up to 25 nm. As the film gets thicker than 25 nm, the AHE shows only IMA behavior and no switching is observed. The switching behavior is generally reproducible across many samples and anneal parameters. For example, in the case of the 9 nm thick film shown in Figure \ref{fig1}, annealing at 100 $^\circ$C leads to AHE with positive sign and robust PMA. When this film is annealed at 200 $^\circ$C, the AHE sign becomes negative, the shape of the curve changes and the coercive field becomes larger. After annealing at 300 $^\circ$C, the $R_{xy}$ value shrinks significantly, and although the sign is positive, hump-like features are present suggesting that there are two competing channels with different AHE signs \cite{Kim2020,Wang2020}. When annealed at 400 $^\circ$C, the AHE sign becomes positive, and the coercivity is also enhanced compared to annealing at 100 $^\circ$C.

Rutherford backscattering, x-ray photoelectron spectroscopy and x-ray diffraction show reduction of the films to a Pd-Co alloy state after hydrogenation \cite{Rimal2021}, which explains the emergence of PMA and AHE behavior.  However, unlike traditional PdCo alloys, hydrogenation of PdCoO$_2$ is unique due to intermixing of Pd and Co along with with trace amounts of oxygen and hydrogen and a novel layered structure can also be realized \cite{Rimal2021}. This may be why the PMA is present in thinner films only. In general, the strength and direction of magnetic anisotropy is determined by many factors such as the crystalline environment (magnetocrystalline anisotropy), strain (magnetostriction), interface and shape \cite{Johnson1996,Tudu2017a}. In the case of Pd/Co multilayers, layer thickness and interface have been found to play critical roles in the development of PMA. A recent study of PdCo alloys found that the surface anisotropy contribution is dominant in thinner films, while shape and magnetoelastic anisotropy dominates thicker films, which can be used to change between IMA and PMA \cite{Harumoto2019}. In a similar manner, the differences in the hysteretic behavior across thickness in our films may be due to similar nanostructural changes resulting from the annealing process.

These results suggest that there may be a continuous transformation of AHE across T$_A$: initially a positive AHE appears, changes to negative with higher T$_A$, then changes back to positive with even higher T$_A$. There are some minor changes in the overall squareness of the curves, but the PMA is preserved. The fact that sign change is limited to thinner films suggests a potentially large role played by interfaces. To understand the sign change, we may consider two competing channels corresponding to the two signs. Initially a positive channel is largely responsible for the positive AHE. After extended annealing, the negative channel dominates and results in a net negative AHE. Finally, with more annealing, the same (or different) positive channel results in positive AHE. To understand this, we discuss how AHE varies across measurement temperatures for the different annealing conditions, i.e. across each of the different AHE signs. 


\begin{figure}[ht]
	\centering
	\includegraphics[width=0.9\linewidth]{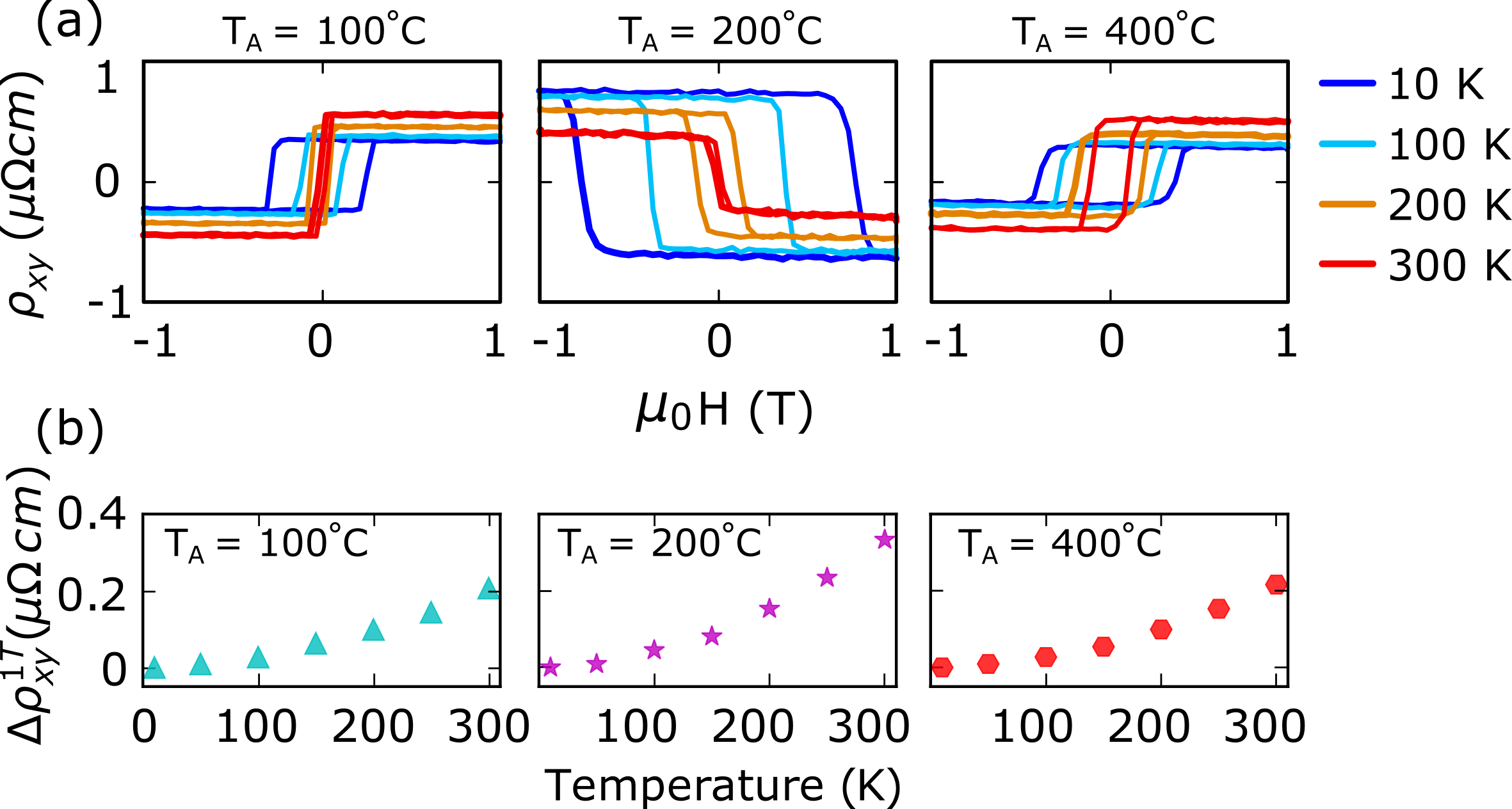}
	\caption{ Behavior of positive and negative AHE for 9 nm thick samples which were annealed at different temperatures. (a) The AHE for 3 different anneal temperatures show different response with temperature. Besides the obvious difference in sign, the scaling of AHE with measurement temperature is opposite between positive and negative AHE. (b) $\Delta\rho_{xy}^{1T} = \rho_{xy}(T) - \rho_{xy}(10K)$ at a field of 1 T: despite the seemingly substantial changes of the anomalous Hall effect in (a), the temperature dependent contribution remains, in fact, little changed throughout these different annealing conditions.  }
	\label{fig2}
\end{figure}

In Figure \ref{fig2}(a), we show the variation in positive and negative AHE across measurement temperatures in 9 nm thick samples. At the lowest anneal temperature of 100 $^\circ$C, AHE is positive and $|\rho_{xy}|$ increases with temperature. At T$_A$ = 200 $^\circ$C, AHE is negative, but $|\rho_{xy}|$ decreases with temperature, which is opposite to the previous case. With an even higher anneal temperature of 400 $^\circ$C, the AHE switches to positive and $|\rho_{xy}|$ increases with temperature. When $\rho_{xy}$ values for each T$_A$ are compared to the lowest measured temperature, as shown in Figure \ref{fig2}(b), we find that at a field of 1 T, where $\rho_{xy}$ is saturated, samples at all anneal temperatures show similar response. This quantity, which we define as $\Delta\rho_{xy}^{1T} = \rho_{xy}^{1T}(T) - \rho_{xy}^{1T}(10K)$, shows that although the overall sign of AHE is different between the different anneal conditions, temperature-dependent contribution to AHE remains little changed. This suggest that there are at least two distinct mechanisms responsible for the observed AHE. One is the temperature-independent part, whose sign depends strongly on the annealing parameter, and the other is the temperature-dependent contribution, which always remains positive and grows with temperature, and is little dependent on the annealing conditions. It should be noted that the Curie temperature (T$_C$) of these films are beyond room temperature \cite{Rimal2021}, and the appearance of two contributions may be a result of two different scattering channels.

Sign change in AHE was previously reported in a few materials including elemental metals \cite{Miyasato2007,Tian2009a} and multilayers \cite{Zhang2016}, but its origin is not always clear. Sometimes, such a sign change can occur when the dominant contribution for AHE changes between extrinsic and intrinsic mechanism \cite{Onoda2006}. Disorder can also strongly influence carrier scattering and lead to changes in AHE \cite{Sinitsyn2005,Guo2014,Bianco2014}, which includes examples such as the metallic oxide SrRuO$_3$ \cite{Kim2020,Wang2020,Wu2020,Skoropata2021}, intermetallic alloys such as MnGa \cite{Zhu2014} and magnetic topological insulators \cite{Zhang2020,Wang2021}. Considering that the bands in hydrogen-reduced PdCoO$_2$ are derived from Pd and Co states, and since the film structures are intermediate between single-crystalline and polycrystalline, the sign change of AHE is likely a combined result of both extrinsic (scattering) and intrinsic (Berry curvature) mechanisms. The resistivity of our films lie close to the region  where crossover between extrinsic and intrinsic mechanisms occurs \cite{Miyasato2007}, providing another hint that the two competing channels may be due to different mechanisms. Also, the fact that thickness plays a role shows that the interfaces may have an important contribution and, as evidenced from Figure \ref{fig1}, thickness and PMA may also be correlated and the sign change may be dependent on the PMA. Yet, considering the structural complexity of these films, it will be difficult to pinpoint the exact origin of the sign change. Regardless, the very fact that the desired AHE sign can be reproducibly achieved with a simple annealing procedure suggests that this effect may be harnessed in this material for further studies on AHE and/or applications. Similarly, it would be interesting to examine the extrapolation of these results to similar systems.

\begin{figure}[ht]
	\centering
	\includegraphics[width=0.9\linewidth]{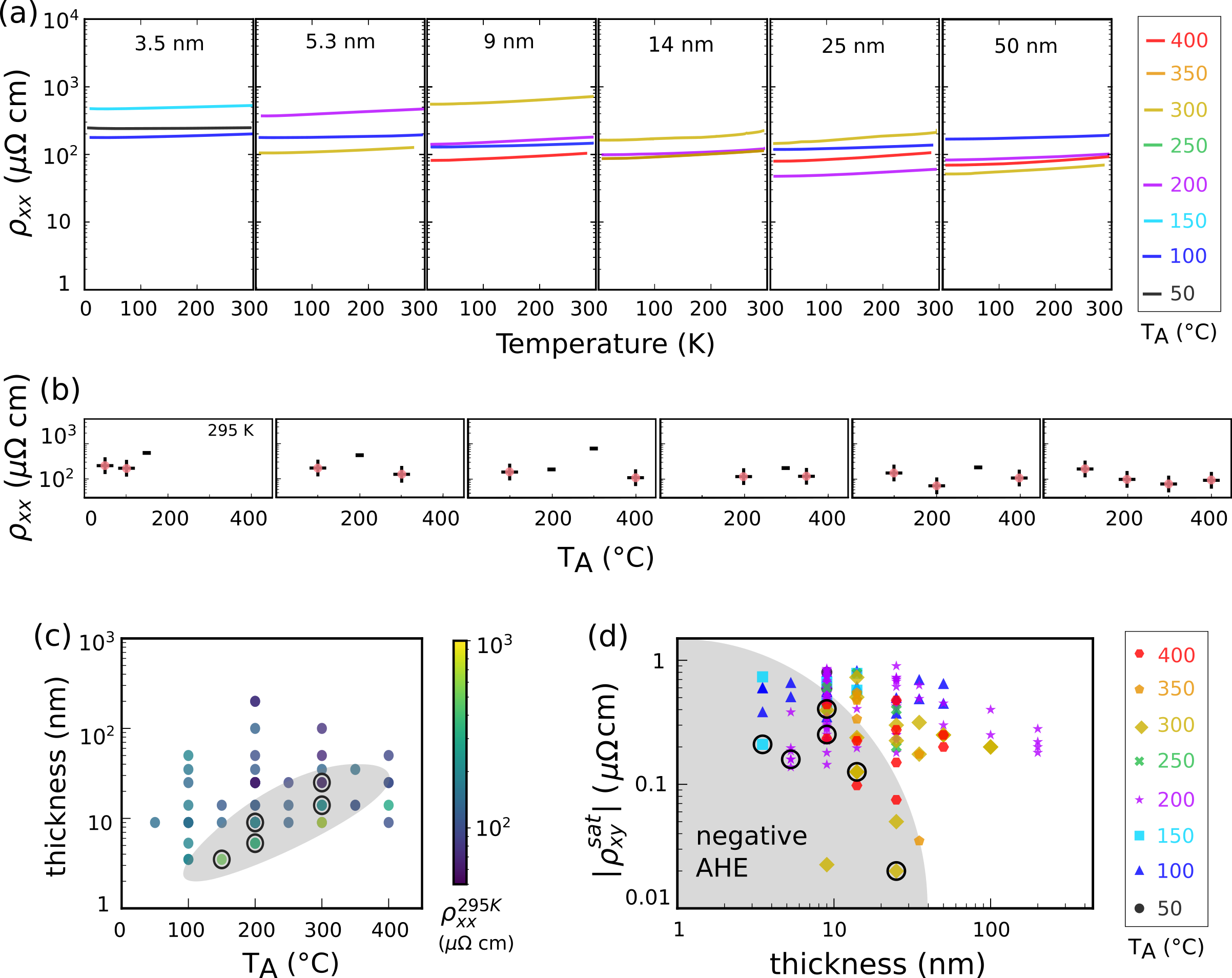}
	\caption{Transport properties of hydrogen-reduced PdCoO$_2$ with different thickness. (a) Temperature dependence of longitudinal resistivity for films of different thickness annealed at different temperatures. (b) $\rho_{xx}$ at 295 K with the corresponding AHE signs across anneal temperature and film thickness. (c) Thickness and anneal temperature dependence of room-temperature resistivity. The resistivity varies by about two orders of magnitude depending on thickness and anneal conditions. The open circles show negative AHE and the shaded region (not drawn to scale) corresponds to region where negative AHE may exist. (d) The variation of saturated value of $\rho_{xy}$ at 295 K with thickness. Markers correspond to different anneal temperatures, with open circles representing negative AHE. The shaded region is loosely drawn to represent a possible region where negative AHE may exist. }
\label{fig3}
\end{figure}

The longitudinal resistivity ($\rho_{xx}$) values with different T$_A$ for the hydrogen-reduced films are shown in Figure \ref{fig3}(a). In general, the hydrogenated films are more resistive than pristine PdCoO$_2$ \cite{Brahlek2019}. It is also notable that the residual resistivity ratio (RRR), defined as $RRR = \rho(295 K) / \rho(10 K)$, is substantially reduced after hydrogenation. Quantitatively, while RRR in pristine PdCoO$_2$ thin films grows with thickness and reaches about 16 at 100 nm, it is relatively constant at about 1.2 $\pm$ 0.1 for all the hydrogenated films, regardless of thickness. Both increased resistivity and reduced RRR after hydrogenation show that the hydrogenated films have substantially higher level of defect scatterings than the pristine films. Furthermore, as shown in Figure \ref{fig3}(b), $\rho_{xx}$ is usually higher for negative AHE, suggesting that the additional scattering channel responsible for the higher resistivity is responsible for the negative AHE. Figure \ref{fig3}(c,d) compiles the relationship between thickness, anneal temperature and AHE which provides a guide to tune the sign and magnitude of AHE in different films by varying the anneal parameter T$_A$. It is notable that negative AHE can show up only within a narrow region of the phase space.


\section*{Conclusion}

In summary, we showed that below a critical thickness of about 25 nm, magnetic and electronic properties of the hydrogen-reduced PdCoO$_2$ films can be effectively tuned with the annealing temperature. Then, beyond the critical film thickness, PMA with sign tunable AHE changes to IMA with a fixed sign of AHE. Competitions between different scattering mechanisms, coupled with out-of-plane anisotropy, may lead to the AHE sign change.




\section*{Acknowledgements}
This work is supported by National Science Foundation (NSF) Grant No. DMR2004125 and Army Research Office (ARO) Grant No. W911NF2010108. We thank Shriram Ramanathan for helpful discussions.

\bibliographystyle{elsarticle-num} 
\bibliography{main}

\end{document}